\documentclass[preprint,showpacs,groupedaddress,floatfix,nofootinbib]{revtex4}

\usepackage{dcolumn}
\usepackage{graphicx}
\usepackage{epsfig}
\usepackage{hyperref}
\usepackage{graphicx}
\usepackage{dcolumn}
\usepackage{amssymb,eucal}
\usepackage{longtable}
\usepackage{amsmath,amssymb,bm}
\usepackage{mathrsfs}

\newcommand{\mathd}{\mathrm{d}}
\newcommand{\mathi}{\mathrm{i}}
\newcommand{\mathe}{\mathrm{e}}

\begin{document}
\title{Quadratic thermal terms in the deconfined phase from holography}
\author{Fen Zuo\footnote{Email: \textsf{zuofen@hust.edu.cn}}}
\affiliation{School of Physics, Huazhong University of Science and Technology, Wuhan 430074, China}
\author{Yi-Hong Gao\footnote{Email: \textsf{gaoyh@itp.ac.cn}}}
\affiliation{Institute of Theoretical Physics,
Chinese Academy of Sciences, P.O. Box 2735, Beijing 100190, China}

\begin{abstract}
Recent lattice simulation has uncovered many interesting properties of SU(N) gauge theory at finite temperature.
Especially, above the deconfinement phase transition all the thermodynamics quantities acquire significant quadratic contributions in inverse temperature. Such a term is also found to dominate the logarithmic of the renormalized Polyakov loop. Using the Hawking-Page transition in Anti-de Sitter space as an example, we show how such contributions can be naturally generated in the holographic approach.



\end{abstract}


 \maketitle

\section{Introduction}
Color confinement in SU(N) gauge theory is believed to be due to the linearly-increasing potential between colored objects when they are separated. Such a distance dependence is supported by lattice data.
At short distances, the linear potential appears as the quadratic correction to the lead Coulomb term~\cite{Akhoury:1997by}. Such a correction could be imitated through a tachyonic gluon mass~\cite{Chetyrkin:1998yr}. However, the theoretic origin is not clear because no dimension-$2$ gauge invariant operator exists. The linear potential leads to the area law of Wilson loop expectation value. In the gauge/string duality~\cite{Maldacena:1997re,Gubser:1998bc,Witten:1998qj}, the Wilson loop expectation value can be approximately evaluated from the minimum area of the corresponding string worldsheet~\cite{Maldacena:1998im,Rey:1998ik}. Area law for the Wilson loop can be realized when the bulk spacetime ends before the worldsheet goes into the deep infrared region~\cite{Witten:1998zw}. Along this line, the relation between linear potential and quadratic corrections in the two point correlation functions is further studied in~\cite{Andreev:2006ct,Zuo:2008re}.

Now let us extend the discussion to finite temperature. Recent lattice data has uncovered some similar phenomena in the deconfining phase of SU(N) gauge theory. It is pointed out in \cite{Meisinger:2001cq} that early lattice data for the trace anomaly in SU(3) gauge theory~\cite{Boyd:1996bx}, scaled by $T^4$, is dominated by a quadratic term in inverse temperature. With an additional quartic term, the data can be fitted extremely well. Such an behavior can be made manifest when the trace anomaly is plotted scaled by $T^2$~\cite{Pisarski:2006yk}. The result is almost constant in the temperature region $T_c\lesssim T\lesssim 4 T_c$. Similar as the quark potential, the thermal quadratic term could also be induced from the presence of massive gluons~\cite{Meisinger:2001cq} (see also \cite{Alba:2014lda}). Compared to the quartic term related to the bag constant, the quadratic term is suggested to reflect the bag thickness in the relevant temperature region~\cite{Pisarski:2006yk}. The corresponding model with both of them, and even higher power terms, is then called ``Fuzzy Bags". Such a behavior is further confirmed for pure SU(N) gauge theory with various $N$~\cite{Panero:2009tv}, and suggested to hold even in the large-$N$ limit.

Another observation concerns the behavior of the Polyakov loop above phase transition. At finite temperature, the Polyakov loop attains nonzero value in the deconfining phase, and serves as the order parameter of the deconfinement phase transition~\cite{Polyakov:1978vu,tHooft:1977hy}. Interestingly, lattice results~\cite{Kaczmarek:2002mc} for the logarithmic of the Polyakov loop in SU(3) gauge theory is also dominated by $1/T^2$ term~\cite{Megias:2005ve}. Such a behavior is confirmed recently in SU(4) and SU(5) gauge theory~\cite{Mykkanen:2012ri}, and the coefficient of the quadratic term has little dependence on the gauge group.

As proposed in~\cite{Witten:1998qj,Witten:1998zw}, deconfinement of SU$(N)$ gauge theory in the large $N$ limit can be described as the transition between the thermal gas phase and the black hole phase on the gravity side. A simple example is given for ${\mathcal N}=4$ super Yang-Mills theory on $S^1\times S^3$. Here we use this example to show that in this case, all the thermodynamic quantities, and also the renormalized Polyakov loop, show very similar pattern as those discussed above. In particular, the quadratic contributions are generated consistently in all of them. Previous attempts to generate such terms holographically can be found in \cite{Andreev:2007zv,Andreev:2009zk}. However, the thermodynamic quantities and the Polyakov loop are obtained from different backgrounds. Also in the effective matrix model~\cite{Dumitru:2012fw}, it is difficult to reproduce the shape of the Polyakov loop when the thermodynamic quantities are well fitted.

The paper is organized in the following way. In the next section we review the pure thermal solution and the Schwarzschild black hole solution in Anti-de Sitter~(AdS) space. The thermodynamics and the Polyakov loop result are studied in section III and section IV, respectively. Possible generalization to other spacetime dimension is considered in section V. In the final section we give a short discussion and point out possible generalization of the work.

\section{The black holes in anti-de Sitter space-time}
First we review the black hole solutions given in \cite{Hawking:1982dh}, and further studied in \cite{Witten:1998zw}. As in \cite{Witten:1998zw}, we keep the discussion general for arbitrary spacetime dimension $D=d+2$. Only when interpreting the results as dual to those in the boundary gauge theory, we fix $d=3$. Since only in this case the gravity description is known to be dual to the maximally supersymmetric gauge theory.

The gravity action considered composes of the Einstein term and a negative cosmological constant,
\begin{equation}
{\mathcal I}=-\frac{1}{16\pi G_{d+2}}\int \mathd ^{d+2}y\sqrt{g}(R-2\Lambda),
\end{equation}
where $\Lambda=-d(d+1)/(2L^2)$. As well known, the action admits AdS space solution~\cite{Hawking:1982dh,Witten:1998zw}
\begin{equation}
\mathd s^2=-\left(1+\frac{r^2}{L^2}\right)\mathd t^2+\left(1+\frac{r^2}{L^2}\right)^{-1}\mathd r^2+r^2\mathd \Omega_d^2. \label{eq.TAdS1}
\end{equation}
Here $\mathd \Omega_d^2$ is the metric on the unit sphere $S^d$, and the time coordinate $t$ is periodically identified with period $2\pi L$. The corresponding thermal phase is obtained in the Euclidean formalism with the substitution $\tau=\mathi t$.

The action also admits the AdS-Schwarzschild black hole solution~\cite{Hawking:1982dh,Witten:1998zw}
\begin{equation}
\mathd s^2=-\left(1-\frac{\omega_{d+1}M}{r^{d-1}}+\frac{r^2}{L^2}\right)\mathd t^2+\left(1-\frac{\omega_{d+1}M}{r^{d-1}}+\frac{r^2}{L^2}\right)^{-1}\mathd r^2+r^2\mathd \Omega_d^2,\label{eq.AdS-BH1}
\end{equation}
where
\begin{equation}
\omega_{d+1}=\frac{16\pi G_{d+2}}{{\rm{Vol}}(S^d)d},\quad {\rm{Vol}}(S^d)=\frac{2\pi^{(d+1)/2}}{\Gamma((d+1)/2)}.
\end{equation}
 Later we will see $M$ is the expectation value of the energy, and thus the mass of the black hole. In the limit $M\to 0$, the black hole solution reduces to pure AdS (\ref{eq.TAdS1}). $M$ is related to the horizon $r_+$ as
\begin{equation}
M=\frac{1}{\omega_{d+1}}\left[r_+^{d-1}+\frac{r_+^{d+1}}{L^2}\right],
\end{equation}
and the Hawking temperature is given by
\begin{equation}
T_H=\frac{(d+1)r_+^2+(d-1)L^2}{4\pi L^2r_+}.
\end{equation}
 The temperature has a minimum bound
\begin{equation}
T_{\rm{min}}=\frac{\sqrt{d^2-1}}{2\pi L},\label{eq.Tm}
\end{equation}
above which the black hole solution exists.
As shown in ~\cite{Hawking:1982dh,Witten:1998zw}, the thermal AdS solution (\ref{eq.TAdS1}) dominates at low temperature and the black hole phase is more stable at high temperature. The phase transition between them is interpreted as deconfinement in the boundary gauge theory, which takes place at
\begin{equation}
T_c=\frac{d}{2\pi L}>T_{\rm{min}}.\label{eq.Tc}
\end{equation}

Both the above solutions have the topology $S^1\times S^d$ at the boundary. By suitable choice of the transformation function, the radiuses of $S^1$ and $S^d$ can be chosen to be $\beta_H=T_H^{-1}$ and $L$ respectively. Taking the radius of $S^d$ to infinity, the boundary becomes $S^1\times R^d$. Equivalently, one can choose $\beta_H\to 0$, or $M\to \infty$~\cite{Witten:1998zw}.
Absorbing the divergent quantities by redefining the coordinates, one arrives at the metric
\begin{equation}
\mathd s^2 =-\frac{r^2}{L^2}\left[1-\left(\frac{r_0}{r}\right)^{d+1}\right]\mathd t^2 +\left\{\frac{r^2}{L^2}\left[1-\left(\frac{r_0}{r}\right)^{d+1}\right]\right\}^{-1}\mathd r^2+\frac{r^2}{L^2}\mathd \mathbf{x}^2.\label{eq.AdS-BH2}
\end{equation}
Now the horizon is at $r_0$, with $T_H=\frac{d+1}{4\pi L^2} r_0$. The corresponding thermal solution is simply
\begin{equation}
\mathd s^2 =-\frac{r^2}{L^2}\mathd t^2 +\left\{\frac{r^2}{L^2}\right\}^{-1}\mathd r^2+\frac{r^2}{L^2}\mathd \mathbf{x}^2,\label{eq.TAdS2}
\end{equation}
As expected, the black hole phase in this case dominates at any temperature, corresponding to unconfinement of  ${\mathcal N}=4$ super Yang-Mills theory on $S^1\times R^3$.

\section{Thermodynamics and the trace anomaly}
Now let us turn to the thermodynamic quantities. First we consider the case of infinite black hole mass, when the boundary is $S^1\times R^d$.
In this case no other scale than the temperature exists, so the thermodynamic quantities scale with the temperature according to their dimensions
\begin{equation}
s_0=c T_H^d,~~~\epsilon_0=\frac{cd}{d+1} T_H^{d+1},~~p_0=\frac{c}{d+1} T_H^{d+1},
\end{equation}
where
\begin{equation}
c=\frac{L^d}{4G_{d+2}}\left(\frac{4\pi}{d+1}\right)^d.
\end{equation}
In particular, we have
\begin{equation}
c_S^2\equiv\frac{\mathd p_0}{\mathd \epsilon_0}=1/d,~~~\Delta_0\equiv \epsilon_0-d p_0=0,
\end{equation}
As expected, the velocity of sound takes the conformal value and the trace anomaly vanishes. These results are supposed to be valid at large 't Hooft coupling, where the gravity approximation can be trusted. For $d=3$,  perturbative calculation at zero coupling gives $p_{\rm{free}}=4/3 ~p_0$, differing from the above result by a factor of $4/3$~\cite{Gubser:1996de}.

Now we want to see the deviations of these two relations when the boundary space is compact.
The pressure can be derived from the action, which after subtracting that of the thermal AdS space is
\begin{equation}
P=\frac{{\rm{Vol}}(S^d)}{16\pi G_{d+2}}\left[\frac{r_+^{d+1}}{L^2}-r_+^{d-1}\right].
\end{equation}
Then the energy can be derived through the thermal relation
\begin{equation}
E=\frac{d\cdot {\rm{Vol}}(S^d)}{16\pi G_{d+2}}\left[\frac{r_+^{d+1}}{L^2}+r_+^{d-1}\right]=M.
\end{equation}
The entropy can in turn be obtained, which turns out to be exactly the Bekenstein-Hawking entropy of the black hole
\begin{equation}
S=\frac{{\rm{Vol}}(S^d)}{4 G_{d+2}}~r_+^d.
\end{equation}
The corresponding densities can be obtained by dividing the volume of the sphere $S^d$ with radius $L$. For $d=3$, one obtains
\begin{eqnarray}
\epsilon&=&\frac{3N^2}{8\pi^2}\frac{r_+^4+r_+^2 L^2}{L^8}\nonumber\\
p&=&\frac{N^2}{8\pi^2}\frac{r_+^4-r_+^2 L^2}{L^8}\nonumber\\
s&=&\frac{N^2}{2\pi}\frac{r_+^3}{L^6},
\end{eqnarray}
where we have used the duality relation~\cite{Maldacena:1997re}
\begin{equation}
\frac{L^4}{l_s^4}=4\pi g_s N=g_{\rm{YM}}^2 N \equiv \lambda.
\end{equation}
In the high temperature limit, one finds
\begin{equation}
\epsilon\to \epsilon_0,~~~~p\to p_0,~~~s\to s_0,
\end{equation}
which simply reflects the fact that $T_H\to\infty$ is equivalent to making the radius of the boundary $S^d$ infinity.

\begin{figure}[ht]
\centering
	\includegraphics[width=\textwidth]{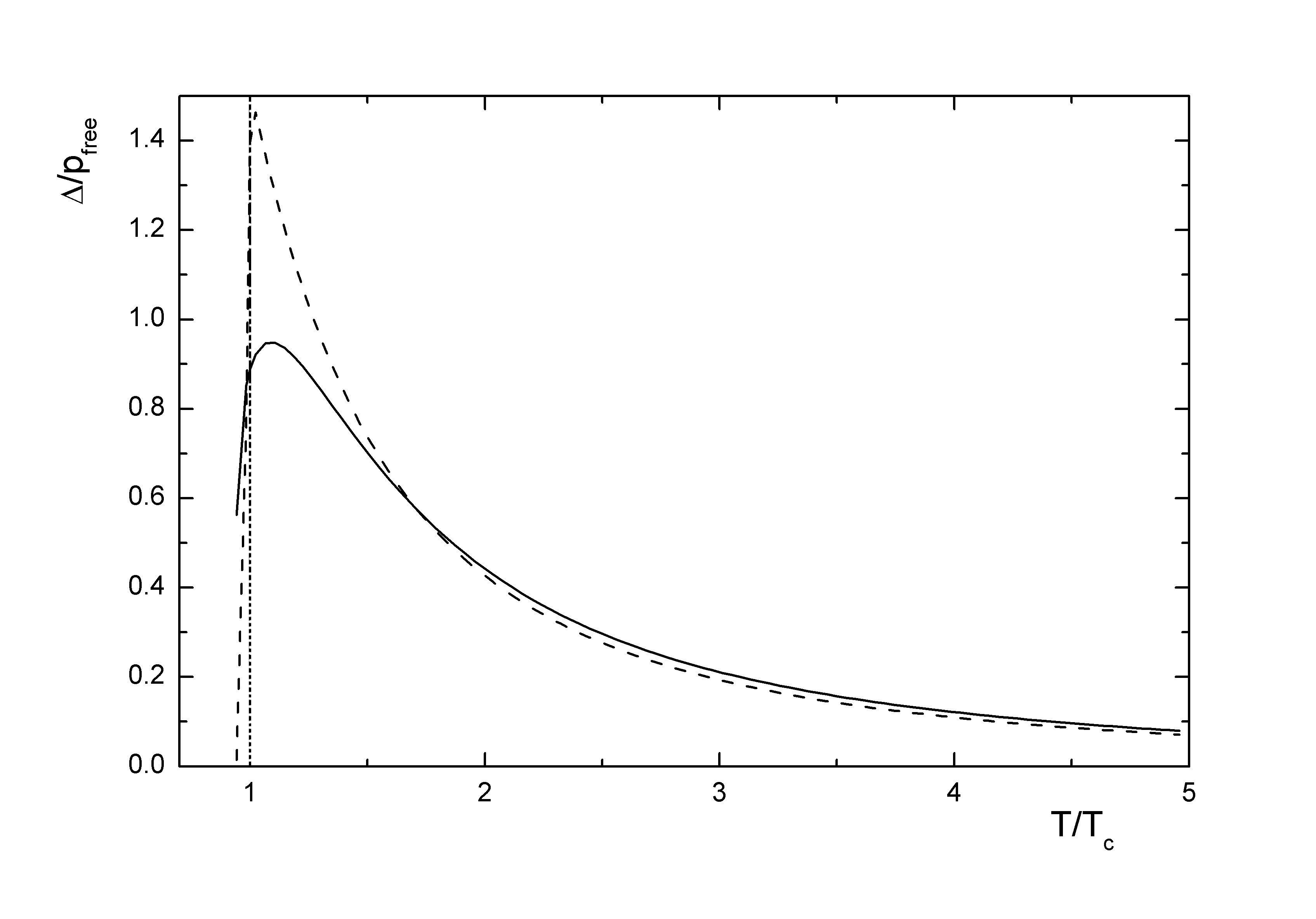}
\caption{\it Holographic result (\ref{eq.D/p}) for the ratio $\Delta/p_{\rm{free}}$ versus $T_H/T_c$ (solid line). Also plotted is the lattice fit (\ref{eq.theta0}) from \cite{Panero:2009tv} (dashed line).}\label{fig:Anomaly}
\end{figure}

The sphere radius introduces another scale into the system. To see the violation of conformal invariance due to this, we formally calculate the quantity
\begin{equation}
\Delta \equiv \epsilon-d p.
\end{equation}
This could be considered as a direct generalization of $\Delta_0$, even though it may not be indeed the trace of the energy momentum tensor when the boundary space is compact. A simple calculation then gives
\begin{eqnarray}
\Delta/p_0&=&\frac{d(d+1)^2}{2^{d+2}\pi^2}\frac{1}{L^2T_H^2}\left[1+\sqrt{1-\frac{T_{\rm{min}}^2}{T_H^2}}\right]^{d-1}\nonumber\\
&=&\frac{d(d+1)^2}{9\cdot2^d}\frac{T_c^2}{T_H^2}\left[1+\sqrt{1-\frac{T_{\rm{min}}^2}{T_H^2}}\right]^{d-1}
\end{eqnarray}
We will mainly focus on $d=3$. In order to compare with the lattice results, we normalize the trace anomaly to $p_{\rm{free}}$ instead
\begin{eqnarray}
\Delta/p_{\rm{free}}&=&\frac{T_c^2}{2~T_H^2}\left[1+\sqrt{1-\frac{8}{9}\frac{T_c^2}{T_H^2}}\right]^2.\label{eq.D/p}
\end{eqnarray}
It turns out $\Delta/p_{\rm{free}}$ exhibits an interesting temperature dependence, as shown in FIG.~\ref{fig:Anomaly}. It increases to a maximum slightly above $T_c$, and starts to decrease with increasing temperature. In the large temperature limit, it goes to zero. The same temperature pattern is found in lattice simulation for pure SU(N) gauge theory.  When $N$ varies, the lattice data can be well fitted by the expression~\cite{Panero:2009tv}
\begin{equation}
\frac{\Delta}{p_{\rm{free}}} =   \left(             
1 - \frac{1}{\left\{
1 + \exp \left[ \frac{(T/T_c)-f_1}{f_2}
\right] \right\}^2}
\right) \left( f_3\frac{T_c^2}{T^2} + f_4\frac{T_c^4}{T^4}
\right),
\label{eq.theta0}
\end{equation}
where $p_{\rm{free}}=\frac{\pi^2}{45} (N_C^2-1) T^4$ is the corresponding perturbative result at zero coupling. The parameters have smooth limits when extrapolated to large $N$. The central values of the fitted parameters at large $N$ are given by
\begin{eqnarray}
f_1&=&0.9918,~\quad  f_2=0.0090, \nonumber\\
f_3&=&1.768,~~\quad   f_4=-0.244.
\end{eqnarray}
In FIG.~\ref{fig:Anomaly} we also plot the lattice fit (\ref{eq.theta0}) with the above parameters. Although the shape of the two curves is similar, the maximum values differ from each other. That also means, the latent heat is different. From (\ref{eq.D/p}) one reads
\begin{equation}
\epsilon(T_c)/p_{\rm{free}}(T_c)=8/9,
\end{equation}
while the lattice fit gives a value around $1.4$.

\begin{figure}[ht]
\centering
	\includegraphics[width=0.5\textwidth]{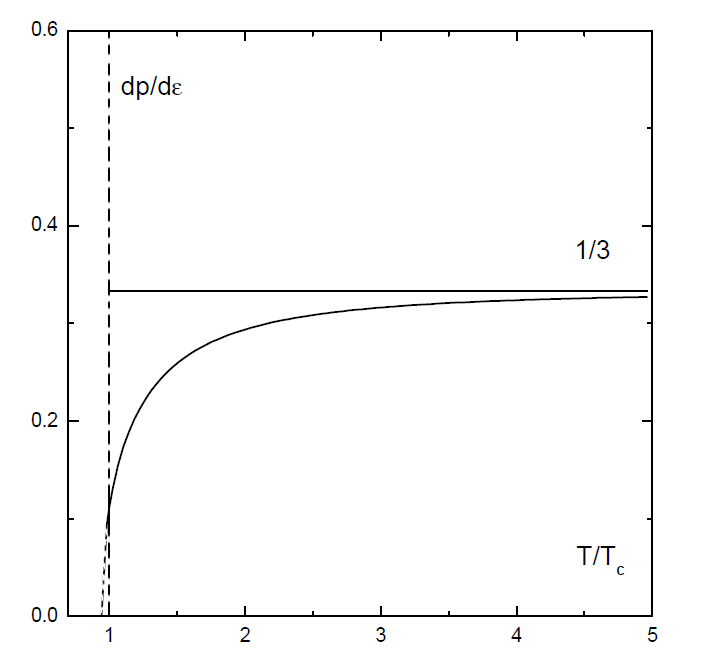}\includegraphics[width=0.5\textwidth]{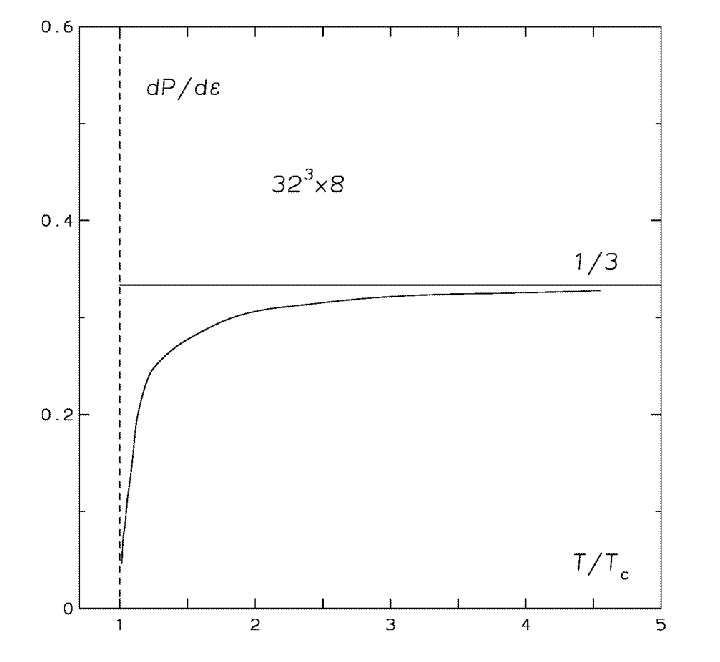}
\caption{\it Holographic result for the squared speed of sound versus $T_H/T_c$ (left), in comparison with the lattice data in pure SU(3) gauge theory \cite{Boyd:1996bx} (right).}\label{fig:v2}
\end{figure}

The appearance of the maximum in the present model can be traced back to two facts. First, the temperature derivative of $\Delta$, and also $\Delta/p_{\rm{free}}$,  goes to $+\infty$ when the temperature approaches its minimum from above. Second, $\Delta/p_{\rm{free}}\to 0^+$ when $T_H\to \infty$. With these two properties, one can generally prove the existence of the maximum. Due to the quickly increasing near $T_{\rm{min}}$, the maximum is expected to appear close to $T_{\rm{min}}$. That means, close to $T_c$. In fact, the shape of all the physical quantities can be obtained by smoothly connecting their asymptotic form in the two limits. There will be no special structure in between, since there is no special point in the black hole solution. In particular, the phase transition is not an intrinsic property of the black hole solution. For example, the ratio $\Delta/s^{(d+1)/d}$ decreases monotonously all the way from $T_{\rm{min}}$ to infinity. Also we can easily sketch the behavior of the squared speed of sound. It goes to $0^+$ when approaching $T_{\rm{min}}$ from above, and increases monotonously to its conformal value $1/3$ in the high temperature limit. Since $T_c$ is close to $T_{\rm{min}}$, the velocity would be slightly above zero at the phase transition.  Explicitly, the squared velocity of the sound is
\begin{equation}
c_S^2\equiv \frac{\mathd p}{\mathd \epsilon}=\frac{1}{d}\sqrt{1-\frac{T_{\rm{min}}^2}{T_H^2}},
\end{equation}
and takes the value $1/d^2$ at $T_c$. The behavior of the velocity square at $d=3$ is plotted in FIG.~\ref{fig:v2}, which turns out to be extremely similar to the corresponding lattice result for pure SU(3) gauge theory~\cite{Boyd:1996bx}.
Similar description for the thermodynamic quantities have already been constructed in the gravity-dilaton system~\cite{Gubser:2008ny,Gursoy:2008bu,Gursoy:2008za}. It has also been extended to describe the behavior of the bulk viscosity~\cite{Gubser:2008yx,Gursoy:2010fj} and the Polyakov loop~\cite{Noronha:2009ud,Noronha:2010hb}. In a similar way, one may explain the enhancement of jet quenching around phase transition, recently pointed out in \cite{Li2014}.



We want to compare our result for the trace anomaly with the lattice fit in more detail.
Above the phase transition, the lattice fit (\ref{eq.theta0}) simplifies due to the large exponential factor as
\begin{equation}
\frac{\Delta}{p_{\rm{free}}} \approx \left( f_3\frac{T_c^2}{T^2} + f_4\frac{T_c^4}{T^4}\right).\label{eq.theta1}
\end{equation}
Such a fit form was first proposed in \cite{Meisinger:2001cq}. It means that above $T_c$ the trace anomaly is dominated by a $T^2$ term, with a small constant correction. Due to the smallness of $f_4$, a simplified model, with only the quadratic term, is considered~\cite{Pisarski:2006yk}. The corresponding pressure density is then
\begin{equation}
 p/p_{\rm{free}}=1-\frac{f_3}{2}\frac{T_c^2}{T^2},
 \end{equation}
where it has been assumed $p\to p_{\rm{free}}$ as $T\to \infty$. For pure glue, $p(T_c)$ is argued to be small, which leads to the estimation $f_3\simeq 2$. In general a series of terms in inverse temperature square can be included. The pressure then takes the form
 \begin{equation}
 p/p_{\rm{free}}=1-\frac{f_3}{2}\frac{T_c^2}{T^2}-\frac{f_4}{4}\frac{T_c^4}{T^4}+...
 \end{equation}
which is called the fuzzy bag model~\cite{Pisarski:2006yk}.

Taking the high temperature expansion of (\ref{eq.D/p}), we find
\begin{equation}
\Delta/p_{\rm{free}}
\approx2\frac{T_c^2}{T_H^2}-\frac{8}{9}\frac{T_c^4}{T_H^4}.\label{eq.D/p1}
\end{equation}
Actually such a truncated expansion is almost exact above $T_c$.
Comparing with (\ref{eq.theta1}), we immediately obtain
\begin{equation}
f_3=2,~~~~~~f_4=-\frac{8}{9}. \label{eq.f3f4}
\end{equation}
We see indeed violation of conformal invariance is mainly due to a quadratic correction, with the coefficient $f_3$ close to the lattice value extrapolated to large $N$. The sign of $f_4$ is also the same as the lattice result for pure gauge theory, and differs from that in QCD at zero temperature and at finite temperature \cite{Bazavov:2009zn}. 
However, the absolute value of $f_4$ is about two times bigger than that from the lattice fit. Such a large negative $f_4$ has significant effects around the phase transition, and leads to the smallness of the latent heat.


\section{Renormalized Polyakov loop}
Now we will discuss the results for the Polyakov loop ${\mathcal L}$. In the dual description, ${\mathcal L}$ is approximately evaluated through the minimum area $S$ of the string worldsheet ending on the temporal loop~\cite{Maldacena:1998im,Rey:1998ik}
\begin{equation}
{\mathcal L}\approx \mathe ^{-S}.
\end{equation}
Due to the different topologies of the solutions shown in the previous section, such a string worldsheet may exist or not~\cite{Witten:1998zw}.
The topology of the thermal AdS is $S^1\times S^{d+1}$, while the AdS-Schwarzschild black hole is $R^2\times S^d$. Due to this, in the first case no such string worldsheet exists and the corresponding Polyakov loop is vanishing. The Polyakov loop ${\mathcal L}$ is related to the quark free energy as
\begin{equation}
{\mathcal L}=\mathe^{-F_Q(T)/T}.
\end{equation}
Thus vanishing of ${\mathcal L}$ corresponds to infinite self energy of single quark, and induces confinement. In the latter case, the whole part $R^2$ is just the string worldsheet that ends on the boundary $S^1$. A nonzero value of the Polyakov loop is thus generated, signaling unconfinement in the black hole phase.

Now we calculate the quark free energy explicitly, through the area of the string worldsheet
\begin{equation}
F_Q(T)\approx TS.
\end{equation}
The area is simply that of the $R^2$ part spanned by $t$ and $r$
\begin{equation}
S_1=\frac{1}{2\pi\alpha'}\int_0^{\beta_H} \mathd t \int_{r_+}^\infty \mathd r.
\end{equation}
The integral is divergent. A finite free energy, and thus $L$, can be obtained by slightly shifting the upper bound of the $r$ integral away from the boundary point $r_+\to \infty$. This corresponds to the renormalization procedure. Here we will define the renormalized quark free energy in a different way, by subtracting the corresponding result in the limit of infinite black hole mass. As explained in the previous sections, in this case the black hole phase always dominates over the thermal AdS solution. For $d=3$ this corresponds to the fact that ${\mathcal N}=4$ Super Yang-Mills theory on the boundary $S^1\times R^3$ is always unconfining. Therefore with such a subtraction we can extract the dynamical part of the quark free energy that is related to the phase transition. Similar procedure has been employed in \cite{Andreev:2009zk}.
The corresponding worldsheet area in the large mass limit is given by
\begin{equation}
S_0=\frac{1}{2\pi\alpha'}\int_0^{\beta_H} \mathd t \int_{r_0}^\infty \mathd r.
\end{equation}
The renormalized quark free energy can then be obtained as
\begin{eqnarray}
F_Q^R&\approx& T_H(S_1-S_0)\nonumber\\
&=&\frac{T_H}{2\pi\alpha'}\int_0^{\beta_H} \mathd t \int_{r_+}^{r_0} \mathd r   \nonumber\\
 &=&\frac{\sqrt{\lambda}}{d+1}\left[T_H-\left(T_H^2-T_{\rm{min}}^2\right)^{1/2}\right].
\end{eqnarray}
As a result, the logarithmic of the Polyakov loop is given by
\begin{eqnarray}
-2\log {\mathcal L}^R&=& 2 F_Q^R(T_H)/T_H\nonumber\\
&\approx& \frac{2\sqrt{\lambda}}{d+1}\left[1-\left(1-\frac{T_{\rm{min}}^2}{T_H^2}\right)^{1/2}\right]\label{eq.log-L1}\\
&=&\frac{\sqrt{\lambda}}{d+1}\frac{T_{\rm{min}}^2}{T_H^2}+{\mathcal O}(T_H^{-4}). \label{eq.log-L2}
\end{eqnarray}

It turns out that the leading term, quadratic in inverse temperature, gives the dominant contribution in the whole deconfining phase. The contributions of all the other terms can only be seen around $T_c$.
This is indeed the behavior found in the lattice simulation for pure gauge theory with gauge group SU(3), SU(4) and SU(5)~\cite{Megias:2005ve,Mykkanen:2012ri}, and even in QCD with unquenched quarks~\cite{Megias:2005ve}. The lattice data can be well fitted by the formula
\begin{equation}
-2\log {\mathcal L^R}=  a+b\left(
\frac{T_c}{T}\right)^2. \label{eq.log-L-fit}
\end{equation}
The values of $a$ and $b$ change with the gauge groups, and for SU(3), SU(4) and SU(5) are roughly in the follwoing intervals~\cite{Megias:2005ve,Megias:2007pq,Mykkanen:2012ri}
\begin{equation}
a\sim -(0.1-0.3),~~~~b\sim 1.1-1.8.
\end{equation}
As in the fuzzy bag model, one may estimate the values of $a$ and $b$ if the fit is supposed to be valid in the whole deconfining region. In the high temperature limit ${\mathcal L^R}\to 1$, which forces $a$ to vanish. Close to $T_c$, the renormalized Polyakov loop is found to approach $1/2$ for different gauge groups~\cite{Mykkanen:2012ri}. This fixes $b\approx 2\log 2\approx 1.39$, which is indeed in the above region.

With the expressions (\ref{eq.Tm}, \ref{eq.Tc}), one obtains from (\ref{eq.log-L2})
\begin{equation}
a=0,~~~~~~~~\quad b=\frac{d-1}{d^2}\sqrt{\lambda}.\label{eq.b-Theory}
\end{equation}
For $d=3$, the above lattice result for $b$ then corresponds to the 't Hooft coupling interval
\begin{equation}
\lambda\sim 24-66.
\end{equation}
One finds that the coupling $\lambda$ is indeed large. In other words, it is possible to reproduce the lattice result for $b$ with a reasonable coupling in a self-consistent way. This can be compared with the calculation of the jet quenching parameter using gauge/string duality~\cite{Liu:2006ug}. Choosing $\lambda=6\pi$, the result for the jet quenching parameter is in agreement with that determined from RHIC data. One may also confirm this with the value of ${\mathcal L^R}$ at $T_c$. From (\ref{eq.log-L1}) one finds
\begin{equation}
{\mathcal L^R}(T_c)=\exp\left[-\frac{d-1}{d(d+1)}\sqrt{\lambda}\right].
\end{equation}
Requiring ${\mathcal L^R}(T_c)\approx 1/2$ for $d=3$ gives
\begin{equation}
  \lambda\approx 36\log^2 2\approx 17.3,~~~b\approx 4/3~ \log 2 \approx 0.9. \label{eq.lambda}
\end{equation}
\begin{figure}[ht]
\centering
	\includegraphics[width=\textwidth]{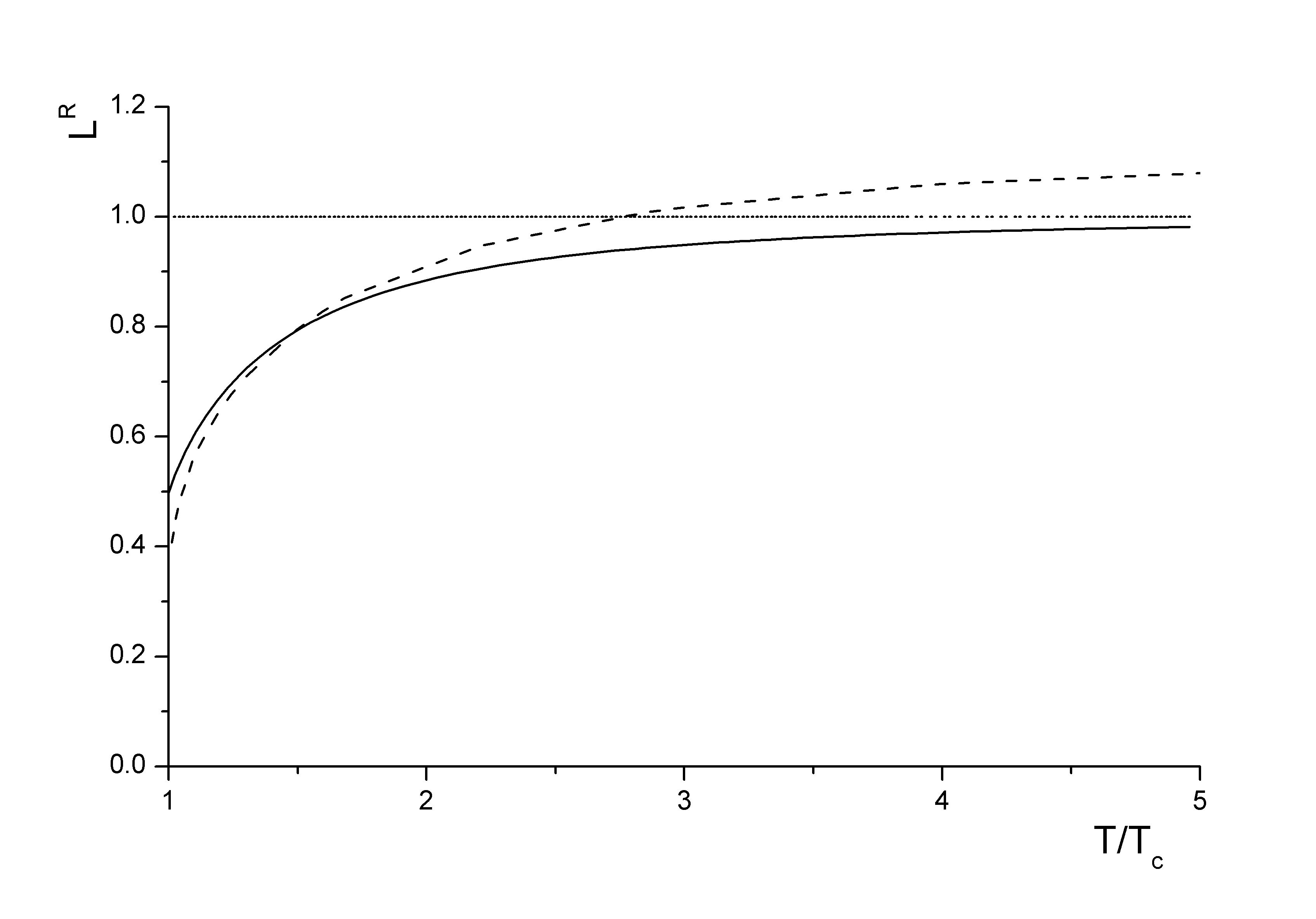}
\caption{\it Holographic result (\ref{eq.log-L1}) with $d=3$ and $\lambda$ given in (\ref{eq.lambda}) for the renormalized Polyakov loop versus $T/T_c$ (solid line), in comparison to the lattice data for pure SU(3) gauge theory~(dashed line)~\cite{Gupta:2007ax}. The asymptotic value ${\mathcal L^R}=1$ is also plotted (dotted line).}\label{fig:Polyakov}
\end{figure}
The result for the 't Hooft coupling$\lambda$ is a little smaller than directly by fitting the parameter $b$. This is of course due to the deviation of our prediction for $a$. In FIG.~\ref{fig:Polyakov} we plot our result for the renormalized Polyakov loop with such a 't Hooft coupling, and compare it to the lattice result for pure SU(3) gauge theory. Notice that the early data here shows the Polyakov loop is slightly smaller than $1/2$ at $T_c$. Above $3~T_c$ the deviation between the two curves becomes large, where the lattice result overshoots the asymptotic value. Again this is due to the difference between our prediction and the lattice fit (\ref{eq.log-L-fit}) for the parameter $a$. Introducing a nonzero value of $a$ as in \cite{Andreev:2009zk} by hand reduces the deviation. However, it is not clear if the fithe deviation could be due to the perturbative contributions not included here. As discussed in ~\cite{Megias:2005ve,Mykkanen:2012ri}, the perturbative result approaches the asymptotic value from above, which could be the reason why the Polyakov loop overshoots unity.

\section{Results for lower spacetime dimension}
The lattice data in $2+1$ dimension shows that the trace anomaly, normalized in unit of $T^3$, is dominated by a linear term in inverse temperature~\cite{Bialas:2008rk,Caselle:2011mn}. This could be compared with the behavior of the quark potential at zero temperature. We know in this case the perturbative potential is logarithmic, and the linear confining potential, if existing, would be roughly the linear correction at short distance. It is also proposed that such a term could be completely due to the leading perturbative corrections~\cite{Bicudo:2014cra}. Our present calculation shows that for any space dimension $d\ge1$, the quadratic term in inverse temperature makes the dominant contribution to $\Delta/T^d$. The contradiction with the lattice result indicates that the Hawking-Page transition is not proper to describe deconfinement in $2+1$ dimension. 
This can be anticipated, since we do not know whether the duality can be generalized to $d\ne3$ or not.

 As for the Polyakov loop, the quadratic term is vanishing when $d=1$. This could be compared with the fact that in $1+1$ dimension the pertubative potential by itself is linear. In $2+1$ dimension, eq.~(\ref{eq.b-Theory}) seems to indicate the existence of a nonzero quadratic term. Again, this may not be the true situation, due to the same reasoning as the trace anomaly.

\newpage
\section{Discussion}
We have shown that the quadratic thermal terms observed in lattice data can be generated naturally on the gravity side.
The reason can be traced back to the difference of the bulk metric when the boundary is compact or not. If one naively compares metric (\ref{eq.TAdS1}) with (\ref{eq.TAdS2}), one finds that in (\ref{eq.TAdS1}) there is an additional constant term in the warp factor. The same pattern can also be found comparing (\ref{eq.AdS-BH1}) with (\ref{eq.AdS-BH2}). According to the dictionary of the duality~\cite{Gubser:1998bc,Witten:1998qj}, such a term would correspond to a dimension-2 operator at the boundary, or a tachyonic field in the bulk. In our calculation no such field is introduced, and the correction is completely due to change of the boundary topology.
The comparison between the thermal properties of the theory on $S^3$ and $R^3$ also shows that such contributions appear together with the phase transition. Therefore they could the embers of confinement above $T_c$, i.e., the remaining effects of the linear potential in the deconfined phase. Combining the confinement argument in~\cite{Witten:1998zw} through the Wilson loop, one may suspect that these quadratic terms are also due to global change of the bulk spacetime, rather than some local effects.

The price to pay for such contributions, and confinement, in the present case is the compactness of the boundary space. Therefore the properties found here may not be directly related to those in ordinary space time. However, the idea behind the model could be employed to describe realistic gauge theory. As we have explained, the shape of all the quantities in the model can be obtained by combining the behavior at $T_{\rm{min}}$ and in the high temperature limit. Therefore, the results presented here could be universal among theories with a minimum temperature of the black hole phase. The numerous studies in the gravity-dilaton system have partially confirmed this. In particular, it is pointed out that the existence of a minimum temperature of the black hole solution is related to confinement~\cite{Gursoy:2008bu,Gursoy:2008za}.
If this is in general true, the results shown here may be extended to all confining theories.


\section*{Acknowledgments}

F. Z. thanks Pietro Colangelo, Floriana Giannuzzi and Stefano Nicotri for helpful discussions. This work is partially supported by the National Natural Science Foundation of China under Grant No. 11135011.


\end{document}